# 90° domain wall relaxation and frequency dependence of the coercive field in the ferroelectric switching process

M. H. Lente,[1] A. Picinin, J. P. Rino, and J. A. Eiras

Universidade Federal de São Carlos, Departamento de Física

São Carlos, S.P., Brazil, CEP 13560-905

The mechanisms involved in the polarization switching process in soft and hard $Pb(Zr_{53}, Ti_{47})O_3$ (PZT) bulk ceramics were investigated through the dependency of the hysteresis loop on the frequency. In order to determine the influence of the defects on the domain switching dynamics the samples were characterized in the virgin state and after a fatigue or a depinning process. The frequency dependence of the polarization revealed a strong relaxation of the 90° domain walls at ~100 Hz. The results also revealed a strong influence of the kind of defect and their distribution into the ferroelectric matrix on the domain switching dynamics, which reflected in the frequency dependence of the coercive field and the percentage of the backswitching. Initially, it was observed that the frequency dependence of the coercive field for the soft and the hard PZT in the virgin state had just one rate of change per decade in all frequency range investigated, which is the standard behavior found in the literature. However, after the fatigue or the depinning process two rates of changes were noticed. Consequently, an evidence of an upper frequency limit for the coercive field changes was found. The percentage of the backswitching and its behavior for the soft PZT was almost independent of the fatigue state in all frequency range investigated. Nevertheless, for the hard PZT an opposite behavior was verified. The reorientation of the domains was modeled as occurring in a viscous medium where several forces such as viscous and restoring forces act on them.

[1] E-mail: mlente@df.ufscar.br

## 1. INTRODUCTION

Ferroelectric materials have been widely employed in the technological industry to produce electromechanical and nonvolatile memory devices.[1,2] Several practical applications are directly related to the switching behavior of the spontaneous polarization. Thus, the interest in these technological uses has encouraged intensive theoretical and experimental studies on the polarization switching process in single crystals,[3] thin films[4] and bulk ceramics.[1,5] It is fundamental to understand the domain switching process in order to improve the performance of many electronic devices, mainly those ones based on the ferroelectric properties.

For single crystals, the polarization switching has been interpreted as nucleation-growth processes of the antiparallel domains, spending a very short time to be completed.[6,7] Nevertheless, due to the high complexity of the polycrystalline structure, the domain reorientation process in bulk ceramics[8,9] and thin films[10,11,12] have been presented much more complex than that observed for the single crystals. It has been interpreted as two successive 90° rotation followed by a 90° domain wall rearrangement instead of a pure 180° flipping.[9,13] The low mobility of the 90° domain walls reflects on the long time necessary to polarize the ferroelectric ceramics in order to reach the best electromechanical properties.[14,15] Additionally, it is well known that the switching process is strongly influenced by the presence of defects and grain boundaries.[16,17,18] For instance, a continuous domain reorientation in ferroelectric ceramics may produce strong changes in their switching behavior. These changes are due to the presence of defects that may induce a fatigue[19,20] or a depinning process.[20,21] Commonly, fatigue and depinning processes have been attributed to the presence of space charges[19,22] and dipolar defects,[20,23] respectively. Consequently, smaller polarization and strain values than those expected theoretically[24] as well as a much longer switching time[9,25] are commonly observed in ferroelectric polycrystals.

It is well known that the ferroelectric hysteresis loop is frequency dependent,[26] reflecting in the coercive field and in the polarization behaviors. It has been widely reported that an increase in the frequency produces a continuous increase in the coercive field. This behavior has been observed in single crystals,[26, 27] bulk ceramics[28, 29] and thin films[30]. However, recently, some experimental results showed that the frequency dependence of the coercive field is strongly dictated by the kind of defect introduced into the crystalline lattice, not necessarily increasing with the increase of the frequency.[31] On the other hand, the possible existence of a switching threshold is one of the most fundamental questions regarding the ferroelectricity. According to the continuum theory, there is a lower limit for the coercive field that represents the intrinsic switching barrier.[32] On the other hand, based on random field model, the switching occurs even under a very small field, since time is allowed for polarization switching by polar cluster nucleation.[29] Recent measurements employing ultrasound method provided strong evidence about the existence of a lower limit[33]. Although some interesting works about the frequency dependence of the ferroelectric properties have been published, they have analyzed the domain switching process considering it a defect-free system. Nevertheless, it is imperative to consider both the influence of defects and their distribution into the ferroelectric matrix on the domain reorientation process in order to obtain a more realistic description of the ferroelectric properties.

The aim of this work is to investigate the polarization switching process in soft and hard $Pb(Zr, Ti)O_3$ bulk ceramics through the frequency dependence of the hysteresis loops. In order to characterize carefully the influence of the kind of defects as well as their distribution into the crystalline lattice on the domain dynamics the measurements were done in samples in their virgin, fatigued and depinned states. A qualitative model to explain the domain dynamics is suggested.

## 2. EXPERIMENT

$Pb(Zr_{53}, Ti_{47})O_3$ ceramics were prepared by conventional oxide mixing process and doped with either 1 wt.% $Nb_2O_5$ (softer or donors) or $Fe_2O_3$ (hardener or acceptors), hereafter labeled as PZTN and PZTF, respectively. The precursor oxides were mixed in ball mill, dried and calcined at 850 °C for 3.5 h. Discs shaped samples were sintered at 1250 °C in a saturated PbO atmosphere. The sintered samples with 17 mm in diameter were polished to a thickness of 0.5 mm. After that, they were heat-treated at 600 °C for 30 minutes to release stress induced during the polishing and to remove organic materials. Silver electrodes were painted on both sides of the discs. Scanning electron micrographs showed that the average grain size lies between 3.0 - 3.5 μm.

The frequency dependence of ferroelectric hysteresis loops were characterized for both samples in their virgin state and after a long and continuous domain switching ($\sim 10^5$ cycles) in order to induce a state of either fatigue or depinning for the PZTN and PZTF, respectively.[20,34] The fatigued (PZTN) and depinned (PZTF) states were reached by a continuous polarization switching induced by a bipolar electric field of 25 kV/cm at 1 Hz and at room temperature. The samples were introduced into silicon oil bath and a Sawyer-Tower set up was used for hysteresis loop measurements. A triangular electric field of 25 kV/cm was applied on the samples in a frequency range from 10 mHz to 10 Hz. In order to avoid the self-heating, [20] mainly at high frequencies, only few cycles of the electric field were applied on the samples. All measurements were made at room temperature.

As the $Pb(Zr, Ti)O_3$ ceramics were doped with the same quantity of either $Nb_2O_5$ or $Fe_2O_3$, it is supposed that the same amount of defects were introduced into the samples and that they have similar domain structure. [35] Then, it is possible to consider that the defects introduced in the PZT are embedded in a ferroelectric matrix. [36] Therefore, all changes observed in the polarization switching behavior for both samples, after continuous electric field cycles, may be attributed exclusively to changes in the defects configuration in the samples. [37] Because the virgin state can be recovered through a thermal treatment at temperatures higher than the Curie temperature the microcracking influence on the domain switching process may be neglected. [20]

## 3. RESULTS and DISCUSSIONS

Although the main focus of this paper is not on the study of the fatigue and depinning processes, they will be briefly discussed in order to understand the effect of the redistribution of defects into the ferroelectric matrix, which are induced by continuos polarization switching, on the frequency dependency of the domain dynamics. The fatigue and depinning processes were reported in details in previous works. [20, 34]

### 3.1 Fatigue and depinning processes

Figure 1 (a) shows the representative hysteresis loop results for the PZTN measured in its virgin state at 10 mHz and 1 Hz, whereas figure 1 (b) shows the respective normalized curves ($P/P_S$ vs. $E/E_{Max}$). Analogously, figures 1 (c) and (d) show the hysteresis loop results and the normalized curves, respectively, for the PZTN in the same frequencies but in the fatigued state. [20] Figures 1 (a) and (c) reveal that at lower frequencies the PZTN always reaches both higher polarization and lower coercive field values independently of the fatigue state. However, in relation to the virgin state, the fatigued one shows an increase in the coercive field as well as a remarkable reduction in the polarization values, as observed in figure 1 (c). It has been proposed that successive domain reorientation in PZT ceramics modified with softener elements (e.g. $Nb_2O_5$) induce a fatigue process promoted by space charges, which are mainly trapped at domain boundaries. [20, 38, 39] Consequently, the domain rotation is hindered reducing gradually the domain degree reorientation and increasing simultaneously the coercive field. [19, 20] It is believed that that space charges are composed mainly by oxygen vacancies. [40, 41]

Analogously, figure 2 (a) shows the representative hysteresis loops data for the PZTF in the virgin state measured at 10 mHz and 1 Hz, whereas figure 2 (b) shows the respective normalized hysteresis loops. In addition, figures 2 (c) and (d) show the hysteresis loop results and normalized curves, respectively, for the PZTF measured in the depinned state, which was also induced after a continuous polarization switching ($\sim 10^5$ cycles). [20] Contrarily to observed for the PZTN, after the continuous polarization switching the PZTF sample presented a remarkable increase in its polarization and coercive field. This fact is related to the presence of the dipolar defects (also called complex defects) that in the virgin state are aligned parallel with the polar direction of domains, [37, 42] acting as pinning agents for the domain motion. [9, 21, 34] Nevertheless, when an external electric field is switched the ferroelectric domains tend to reorient, forcing these defects to realign perpendicularly to the polar direction of domains, thus starting a depinning process of the switchable domains. [37, 34] Then, the polarization values increase due to a higher alignment of domains with the electric field. [20, 34] The data also revealed that for the soft PZT fatigue is the unique process observed during all switching time. However, for the hard PZT, it was found that although fatigue and depinning processes occur simultaneously, the depinning process is dominant in the beginning, while fatigue pass to be notice only at the end of the switching process. [43]

### 3.2 Characterization of the evolution of the domain switching dynamics

In order to investigate the domain reorientation dynamics and the influence of the rearrangement of the defects on this dynamics the frequency dependence of the hysteresis loops were characterized in the virgin state and in the either totally fatigued (PZTN) or totally depinned state (PZTF). An intermediate case is also shown. From these results, the percentage of backswitching [(1 - $P_R/P_S$) x 100%] and the relative coercive field [$E_C/E_{Max}$] were calculated for each frequency from the respective normalized hysteresis loops (figures 1 and 2 (b, d)). The results for the PZTN are shown in figure 3 (a) and (b) and for the PZTF in figures 3 (c) and (d). Figure 3 (a) shows that the backswitching for the PZTN in both states (virgin and fatigued) is strongly frequency dependent only at lower frequencies (< ≈400 mHz) and becomes reasonably frequency independent at higher ones. At lower frequencies there is also a tendency for the virgin state to present a lower backswitching in comparison with the most fatigued one. Additionally, it is noticed that before the fatigue process the coercive field for the PZTN increases gradually with the increase of the frequency up to ≈ 5 Hz (figure 3(b)). Then, after that it tends to saturate. However, after the fatigue, the coercive field increases smoothly until ≈ 400 mHz and after that it tends clearly to be frequency independent. It is verified that in the partially or fully fatigued state the coercive field is higher than in the virgin one in all frequency range investigated.

Figures 3 (c) and (d) show the frequency dependence of the ferroelectric properties for the PZTF. Contrarily to the PZTN, the percentage of the backswitching and its frequency dependence were significantly dependent on the continuous polarization switching process. Initially, in the virgin state, the backswitching was nearly frequency independent reaching almost 62%. Nevertheless, after the depinning process, it passed to be frequency dependent at lower frequencies (< ≈400 mHz) and being independent at higher ones. For the most depinned state the backswitching values were 53% and 23% at lower and higher frequencies, respectively. In addition, figure 3 (d) reveals that the coercive field followed the same tendency of the backswitching. Initially it was frequency independent. However, after the depinning process it became frequency dependent at lower frequencies, increasing slightly its absolute value with the increase of the frequency but tending to be frequency independent at higher ones. The depinned state always presented higher coercive field.

### 3.3 Influence of the rearrangement of the defects on the polarization and backswitching

Theoretically, it is expected that the highest polarization values for ferroelectric ceramics with tetragonal and rhombohedral symmetries are respectively 0.831 and 0.866, in relation to single crystals, since all allowed directions could be completely reversed under high electric fields. [44, 45] Nevertheless, the results in figures 1 and 2 show that the polarization values are much smaller than that predicted theoretically to PZT ceramics. [24] This fact can be understood considering the effects of the microstructure (e. g. grain size and porosity) and defects introduced into the lattice on the domain switching properties. First, the relative lower polarization values may be explained based on the depolarizing fields [18] and intergranular stress associated with the grain boundary [18, 46]. These factors hinder strongly the domain reorientation reducing significantly the total domain switching fraction, [18] thus limiting the polarization process. On the other hand, the backswitching behavior observed in our measurements can be considered an important feature related directly to the interaction between domains and defects. This hypothesis is valid since the microstructure in our measurements is the same for both samples (the same grain size) and unaltered by switching (no microcracking) [20] during all fatigue and depinning processes. Figure 3 (a) shows that, for each frequency, the percentage of backswitching for the PZTN is almost independent of the fatigue state in all frequency range investigated. It means that independently of the configuration of the space charges into the ferroelectric matrix the same percentage of backswitching is always observed. However, for the PZTF, the remarkable interaction between dipolar defects and domains may be systematically identified through the hysteresis loop evolution. Initially, in the virgin state, when all dipolar defects are aligned along to the direction of the spontaneous polarization of each domain, the strong electrical interaction between domains and dipolar defects [37] is responsible for the small realignment between domains with the applied electric field [9, 20]. This high interaction results in very small polarization values and high backswitching. Nevertheless, due to continuous realignment of these defects perpendicularly to the polar direction of domains [37, 34] the interaction domain- dipolar defects is significantly reduced, thus allowing higher polarization values as well as a smaller backswitching. The extreme situation is when all dipolar defects tend to be oriented perpendicularly to the polar direction of domains and, consequently, the interaction domain-dipolar defects tend to become null. In this case, the backswitching is considerably reduced, resulting in a frequency dependence similar to that observed for the PZTN (figures 3 (a) and (c)).

### 3.4 Frequency dependence of the saturation polarization - The relaxation effect

Figure 4 shows the saturation polarization ($P_S$) dependence on the frequency ($\omega$) for the PZTN in the virgin and fatigued states obtained from the hysteresis loops measurements. In addition, it is also showed the results obtained through the current transient measurements, recently reported, [9] converted to the frequency domain. The results show a very good accordance between the results obtained from the hysteresis loop measurements with those obtained from the transient. It is noticed a decay of the polarization with the increase of the frequency for all curves. However, a remarkable relaxation of the polarization for the virgin sample is observed at $\approx$100 Hz. Recently, it was reported that the main contribution to the total polarization in ferroelectric bulk ceramics is due to 90° domain walls motion, which reorientation spends a long time to be completed. [8, 9] Therefore, it is possible to suppose that the mechanism responsible for the relaxation of the polarization at higher frequencies can be attributed to the 90° domain walls. In summary, the polarization switching process observed in figure 4 may be understood as follows. Under the action of high bipolar electric fields 180° domains switch faster but immediately followed by the 90° domain rotation. In the hysteresis loop measurements these processes cannot be distinguished (in the frequency range investigated here). Nevertheless, the slow decrease of the polarization with the increase of the frequency already indicates the gradual decay of the 90° domain wall contribution. Finally, at ~100 Hz the amount of reorientable 90° domain walls decreases, thus occurring the relaxation process. Figure 4 shows that the relaxation of the 90° domain walls reduces sensitively the saturation polarization values, revealing clearly their importance to the total polarization, which corroborates the prediction made in previous studies. [9]

In order to check the assumptions above a simple test was done to confirm the existence of a relaxation mechanism. A virgin PZTN sample having a disc shape was poled by a d.c. field of 25 kV/cm during 30 min. at room temperature. After that, the planar coupling factor ($k_P$) was calculated by the resonance method. [35] The resonance curve is shown in figure 5 (a) and the $k_P$ value in table I. It must be stressed that the field and temperature employed to pole the samples are not high enough to reach the saturation of the piezoelectric proprieties. [14, 15, 47] After the piezoelectric characterization and using the same apparatus employed in the hysteresis loop measurements, a bipolar electric field of 25 kV/cm and 60 Hz was applied on the sample in order to depoling it. In the beginning, obviously the hysteresis loop presented a preferential direction due to the poling process, as shown in the insert in figure 5 (a). After a continuous polarization switching the hysteresis loop became gradually symmetric (similarly as observed in figure 1(a)), which suggest strongly that the sample is not truly polarized. Remarkably, after this *"depoling process"*, a piezoelectric activity was still noticed, as shown in figure 5(b). The planar coupling factor calculated from this resonance curve is presented in table I. The results reveal that even after the supposed depoling process by the ac field the PZTN sample maintained 57% of the original the $k_P$ value. Thus, these results show clearly that some polarization mechanism cannot be reoriented by bipolar electric fields with frequencies higher than 60 Hz. Consequently, a considerable polarization still remained in the sample, being responsible by the relative high piezoelectric activity. Therefore, these data support the theory that the 90° domain walls in PZT bulk ceramics relax at around 100 Hz.

### 3.4.1 Frequency dependence of the polarization - Relationship between fatigue and power-law

The data in figure 4 also reveal that the polarization curve follows a power-law dependence [$P_S(\omega) \propto \omega^{-n}$]. It is verified that the exponent (n) in the power-law is significantly higher for the fatigued sample than that found in the virgin one (n = 0.052 and n = 0.175 for the virgin and fatigued state, respectively). This means that the polarization change rate (dispersion) in the fatigued state is much higher. The literature reports that the propagation properties of elastic waves in crystalline solids and their interaction with defects are also governed by a power-law. [48, 49] It is shown in the dislocation theory that an increase in the concentration of pinning points corresponds theoretically to an increase in the exponent in the power-law. [48] Then, by analogy with the dislocation theory, it is believed that the continuos trapping of the space charges in the 90° domain walls due to the fatigue process, which evidently increases their concentration in the walls, increases significantly the 90° domain wall relaxation. In the other words, in the fatigued state, which correspond to higher values of the exponent *n*, the gradual increase of the frequency quickly excludes the 90° domain walls contribution to the polarization switching due to their pinning or similarly due to their higher inertia. Analogies between ferroelectric domain wall movement and dislocation models have already been pointed out. [29, 50]

### 3.5 Frequency dependence of the backswitching

Excluding the virgin PZTF sample, all measurements showed that the percentage of backswitching for both compositions was always higher at lower frequencies, becoming almost frequency independent at frequencies higher than ≈400 mHz independently of the fatigue or depinning state (figures 3 (a) and (c)). First of all, it must be emphasized that at lower frequencies both samples reached higher polarization values. It is due to the longer time of application of the electric field (low frequency), which allows a higher realignment of domains with the field. In ceramics, to induce the $90^o$ domain reorientation from their equilibrium position high electric fields are necessary due to their high threshold field level, [9, 28] thus inducing a strong internal mechanical stress. [18, 28, 51] Therefore, mechanically this process may be visualized as being composed by a restoring force ($F_R$) that forces the domains to come partially back to their equilibrium position. Additionally, it may be assumed that the switching of the domains occurs in a viscous medium. [20] Thus, when domains are moving a viscous force ($F_V$), which is assumed to be proportional to the domain speed, acts contrarily to the domain motion and must also be considered. The importance of the viscosity has been already considered theoretically and experimentally for explaining some polarization switching properties as switching time [5, 52, 53] and self-heating [20] as well as the dielectric properties [54]. By analogy with mechanical systems [55] figures 6 (a) and (b) show schematically, in a first approximation, the electrical ($F_E$), restoring ($F_R$) and viscous ($F_V$) forces acting on the domains during both the application and the removal of the electric field, respectively. During the removal of the field when domains tend to return to their origin, at least partially, the viscous force hinders the domain motion because the viscous force and the restoring force have opposite directions (figure 6 (b)), thus decreasing the percentage of backswitching. However, at lower frequencies the domains move much slower to their equilibrium position. As a result, $F_V$ may be neglected and, consequently, only $F_R$ and $F_E$ forces act on them. Therefore, at lower frequencies the domains relax considerably more increasing the percentage of backswitching. On other hand, at higher frequencies the electric field is removed faster and the domains tend to come back faster to their equilibrium position. Consequently, due to their higher velocity, the viscous force increases considerably reducing the backswitching. Remarkably, at frequencies higher than 400 mHz the percentage of backswitching is frequency independent. This result may be explained assuming that at frequencies higher than ~400 mHz the resulting force is zero ($F_E + F_V + F_R = 0$) promoting the same percentage of backswitching. Indeed, recently, it was suggested that the domain wall motion is restricted by viscous drag caused by the coupling of the domain wall to acoustic phonons.[56]

## 3.6 Frequency dependence of the coercive field

Figure 3 (b) shows that the coercive field for the virgin PZTN increases with the frequency up to ~ 5 Hz and after that it tends subtly to levels out. Nevertheless, it is verified that after the fatigue process, the coercive field for the PZTN is frequency dependent only at frequencies lower than ~ 400 mHz, reaching a saturation at higher ones. Remarkably, for the virgin PZTF sample, its coercive field was practically frequency independent in all frequency range investigated. However, after the depinning process, it was slight frequency dependent at lower frequencies, tending to level out at higher ones with higher values, similarly to the fatigued PZTN.

It is possible to explain the PZTF results for the coercive field assuming that initially, when the dipolar defects are aligned parallel to polar direction of the domains, there is a strong electrical interaction between them, which reduces significantly the threshold field necessary to reorient the 90° domains. [13, 57] Nevertheless, when these defects are forced to realign perpendicularly to the polar direction of domains [34, 37] the threshold field increases, thus increasing the coercive field, as observed in figure 3 (d). On the other hand, the space charges seem to have an opposite effect. Their continuum migration to the domain boundaries always increases the threshold field. These results suggest that only dipolar defects (complex defects) are able to reduce the threshold field while monopolar defects (space charges) seem to exclusively increase it.

As commented above, we are analyzing the polarization reversal mainly through mechanical considerations. It is supposed that both a restoring force ($F_R$) and a viscous force ($F_V$) act on the domains during their reorientation. Therefore, the coercive field may be roughly interpreted as an effective field necessary to overcome these resistance forces ($F_V$ + $F_R$) during the polarization switching (figure 6 (a)). Thus, at lower frequencies when the velocity of the domains is low the viscous force may be neglected and, consequently, the coercive field is reduced. Nevertheless, at higher frequencies the viscous force increases accordingly, increasing the "apparent" coercive field. Finally, in the frequency range where the coercive field is frequency independent may be interpreted as being a range where the resulting force, which effectively acts on the domains, reaches a saturation.

Doubtless, one of the most interesting results found in this work is the fact that the coercive field can reach a saturation value in the relative short frequency range investigated. It is also observed that its behavior is extremely dependent on the configuration of defects into the ferroelectric matrix. It has been reported in the literature that the coercive field has only the same rate of change per decade in all frequency range studied. [26, 27, 28] This result implies in a unique process for the polarization switching, which results in a unique activation field. [27] These data have been obtained for samples in the virgin state, which means without any rearrangement of defects induced by a previous electric story. In this way, these results are similar to those obtained for the PZTN in the virgin state (see figure 3(b)). Nevertheless, for the samples in the fatigued and depinned states the results reveal two rates of changes in the frequency range investigated. Thus, it is clear that the rearrangement of defects into the ferroelectric matrix induces a change in the domain dynamics, as discussed above, reflecting in a change in the effective coercive field behavior. Therefore, in accordance with the polarization behavior reported in figure 4, it seems that the upper limit found for the coercive field at higher frequencies is related to the relaxational process of the 90° domain walls. Then, it is possible to draw the following picture to explain the influence of defects on the coercive field behavior found in figure 3. When the domains are switched without the influence of defects (virgin sample), which means unpinned or weakly pinned by space charges, part of the 90$^o$ domain walls are able to follow the electric field switching up to high frequencies. However, the pinning of the 90° domain walls by space charges increases their coercivity, thus does not allowing their rearrangement at higher frequencies, which implies in a saturation of the coercive field at frequencies lower than that observed for an unpinned state. Therefore, the results in figure 3 suggest the existence of an upper frequency limit for the coercive field, which is reached when domains are strongly hindered. As the domains in the PZTF are always pinned by the dipolar defects or by space charges (figure 3) it justifies why a saturation is always observed in the frequency dependence of the coercive field. The change in the coercive behavior by pinning or depinning may be related to a change in the set of the forces ($F_E$ + $F_V$ + $F_R$) that acts on the domains. Unfortunately, it was not possible to find if there exist a lower threshold field because at extremely low frequencies the electronic current becomes higher than that related to domain contribution,[9] thus limiting the resolution of our measurements. Further works will be conducted in order to investigate the existence of a possible lower limit for the coercive field in polycrystalline materials.

## 4. CONCLUSION

The domain switching dynamics was investigated in soft (PZTN) and hard (PZTF) Pb($Zr_{53}$, $Ti_{47}$)$O_3$ ceramics through the analysis of hysteresis loops obtained at different frequencies. The polarization switching dependence on the frequency revealed a relaxation of the 90° domain walls at around 100 Hz. The data also showed a strong influence of the kind of defect and their distribution into the ferroelectric matrix on the coercive field and the backswitching behaviors. Initially, the frequency dependence of the coercive field for both compositions in the virgin state presented only one rate of change per decade in all frequency range investigated. Nevertheless, after the fatigue and the depinning process two rates of changes were observed and an evidence of an upper frequency limit for the coercive field was noticed. The results showed that the migration of space charges to the domain walls is responsible for the considerable increase in the effective coercive field, whereas only a slight decrease in the backswitching is verified. On the other hand, the presence of the dipolar defects in the PZTF aligned parallel to the spontaneous polarization reduces significantly the coercive field increasing dramatically the backswitching. However, when these defects are reoriented perpendicularly to the spontaneous polarization the interaction domain-dipolar defects tends to be null and, consequently, the ferroelectric properties of the PZTF tend to behave similarly as those observed for the PZTN. Finally, the frequency dependence of the coercive field seems to be governed by a viscous and a restoring force acting on the domains.

**ACKNOWLEDGMENT**

The authors thank FAPESP, CAPES and CNPq for financial support.

# Figure Captions

FIG 1. (a) representative hysteresis loop results for the PZTN measured at 10 mHz and 1 Hz in their virgin state; (b) the respective normalized hysteresis loop results ($P/P_S$ vs. $E/E_{Max}$); (c) hysteresis loop results measured at 10 mHz and 1 Hz in the fatigued state induced after continuous polarization switching; (d) the respective normalized hysteresis loops corresponding to figure (c).

FIG 2. (a) representative hysteresis loop results for the PZTF measured at 10 mHz and 1 Hz in their virgin state; (b) the respective normalized hysteresis loop results ($P/P_S$ vs. $E/E_{Max}$); (c) hysteresis loop results measured at 10 mHz and 1 Hz in the depinned state induced after continuos polarization switching; (d) respective normalized hysteresis loops corresponding to figure (c).

FIG 3. (a) Percentage of backswitching [($1 - P_R/P_S$) x 100%] for the PZTN for the virgin and for each fatigued state; (b) relative coercive field for the PZTN for virgin and for each fatigued state; (c) percentage of backswitching for the PZTF for virgin and each depinned state; (d) relative coercive field for the PZTF for virgin and each depinned state.

FIG 4. Frequency dependence of the saturation polarization for the PZTN in the virgin and fatigued states obtained from the hysteresis loops measurements. The results obtained from the current transient measurements [9] and converted to the frequency domain are also shown.

FIG 5. Resonance curves obtained through the piezoelectric characterization for the PZTN sample: (a) poled by a dc electric field; (b) "depoled" by a 60 Hz bipolar electric field.

FIG 6. Representation of forces acting on domains during the domain reorientation: (a) during the application of the electric field, (b) during the removal of the field.

Table I: Planar coupling factor ($k_P$) for the PZTN calculated after a poling process and after a "depoling" process by a 60 Hz bipolar electric field.

| Sample state | $k_p$ (%) |
| --- | --- |
| Poled by dc field | 28.0 |
| "Depoled" by ac field | 16.0 |

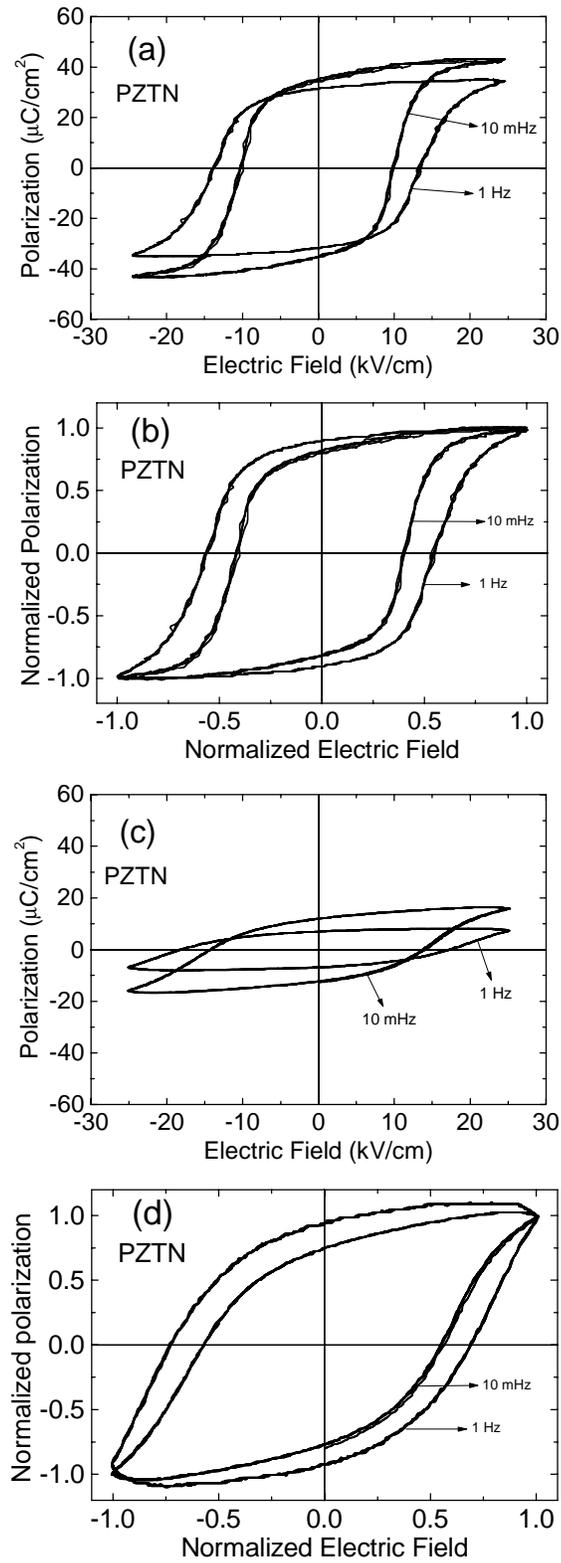

Figure 1: M. H. Lente et al.

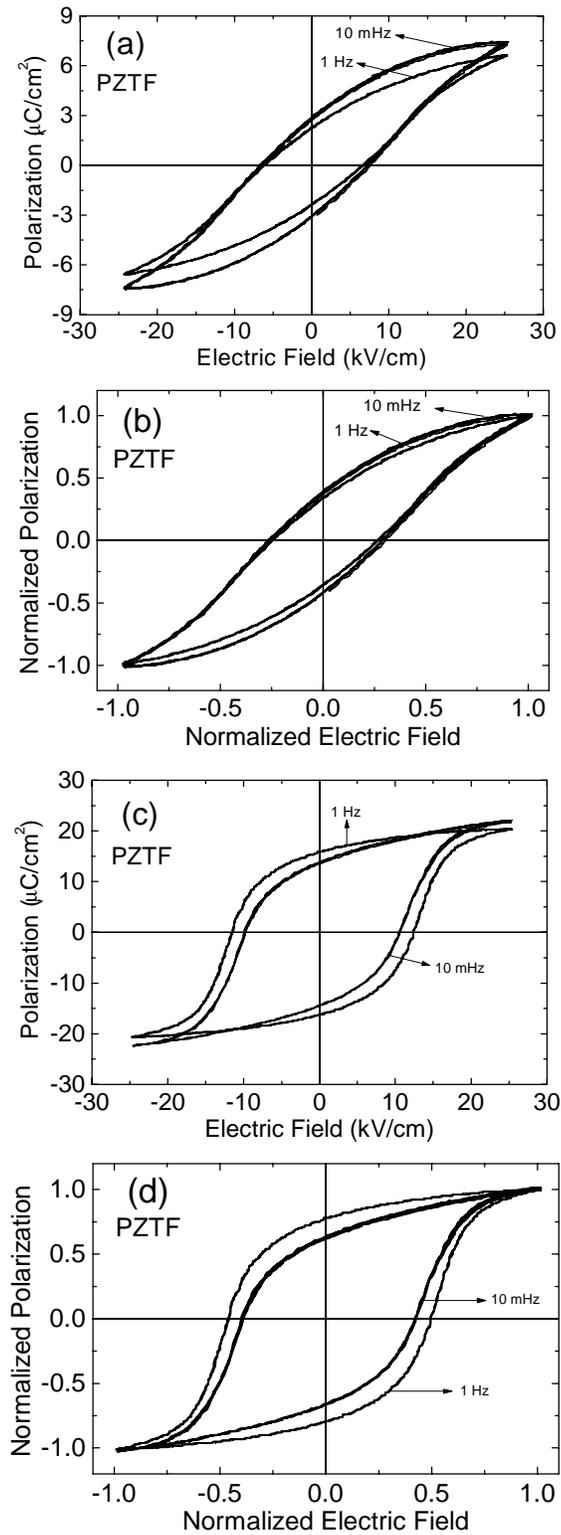

Figure 2: M. H. Lente et al

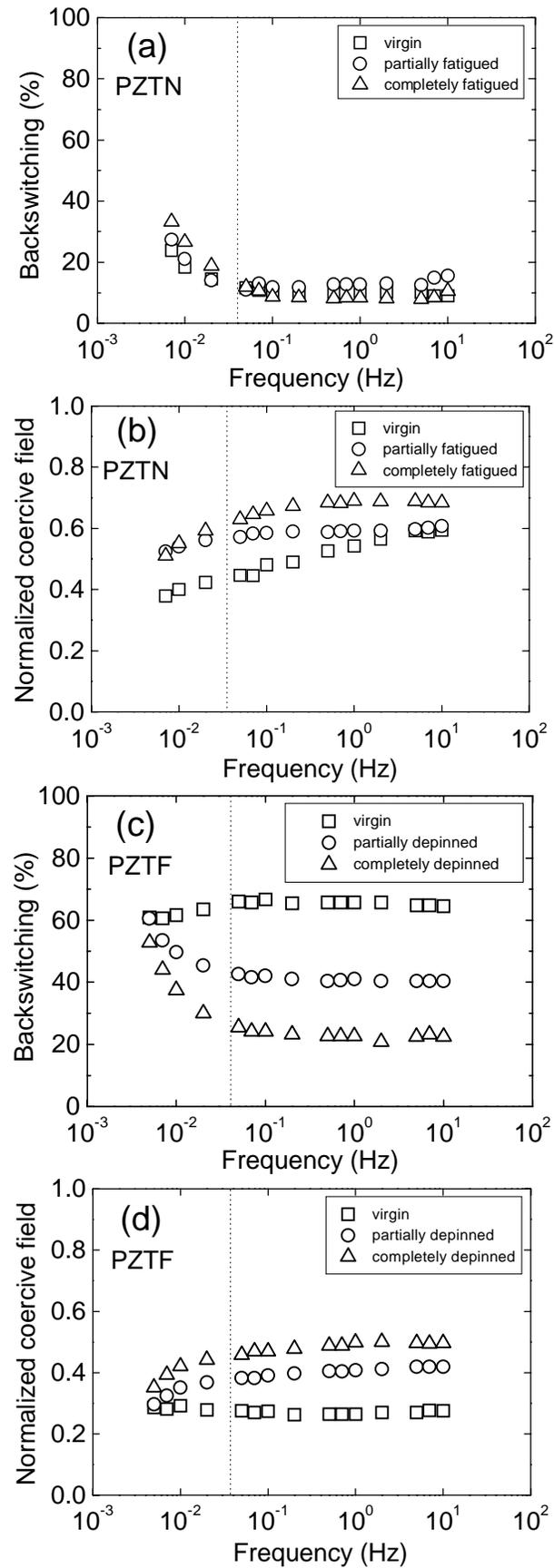

Figure 3 M. H. Lente et al

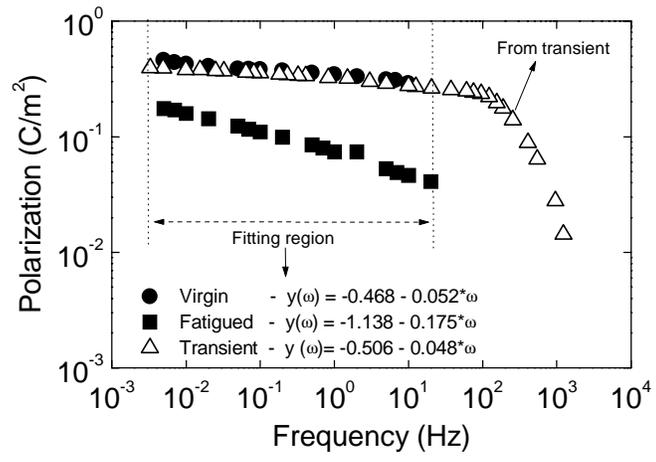

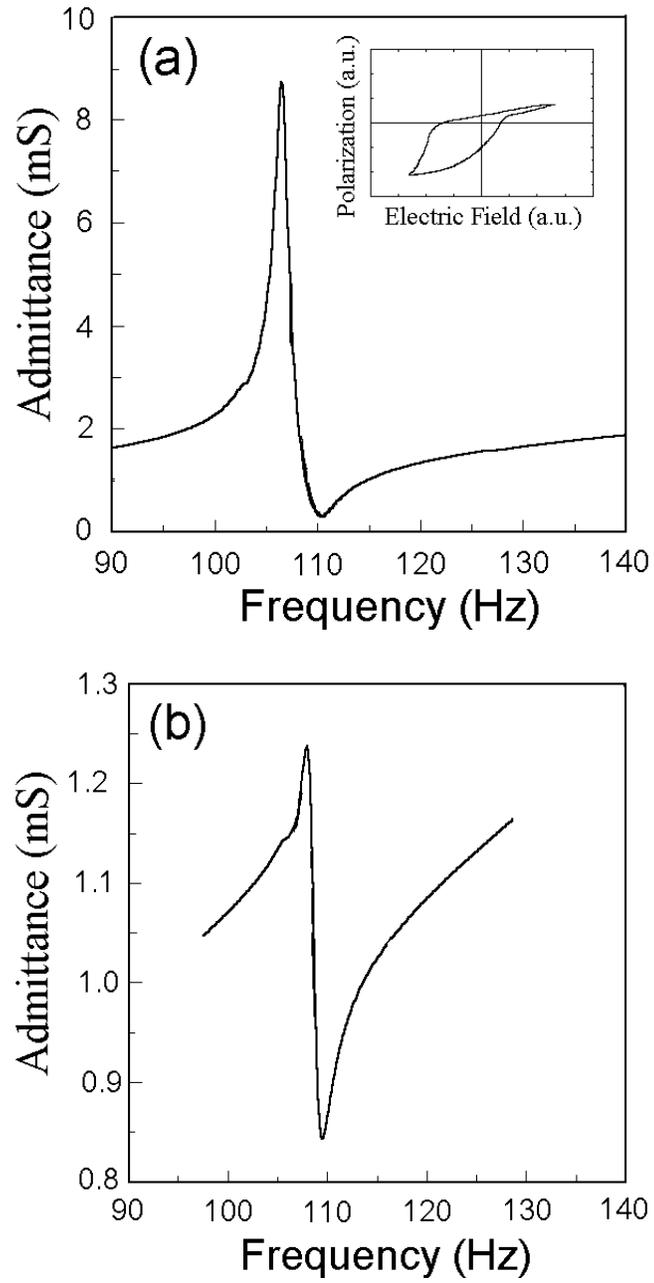

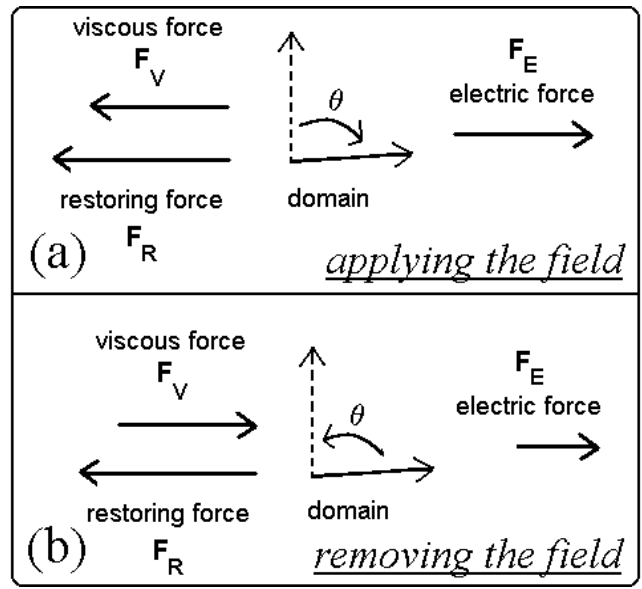

Figure 6: M. H. Lente et al